\documentstyle[10pt]{article}

\author{L.S.F.Olavo\\
Departamento de Fisica, Universidade de Brasilia,\\
70910-900, Brasilia-D.F., Brazil}
\title{Quantum Mechanics as a Classical Theory XVI:\\
Positive-Definite Phase-Space Densities
}

\begin{document}

\maketitle
\begin{abstract}
In this paper we will turn our attention to the problem of obtaining
phase-space probability density functions. We will show that it is possible
to obtain functions which assume only positive values over all its domain of
definition.
\end{abstract}

\vspace{20pt}

\mbox{PACS numbers: 03.65.Bz, 03.65.Ca}

\newpage

\section{Introduction}

Since the very moment, when the phase-space probability density
representation was introduced to reproduce, in a manner as close as possible
to classical statistical mechanics, the results of the expectation values
predicted by quantum mechanics\cite{Wigner1932}, there have appeared the
discomfort of being capable of making this description only appealing to
probability densities having negative values within some range of its
phase-space arguments $(x,p)$.

Many papers have appeared since then trying to push this representation
method to its utmost formal developments\cite{Metha}. The negativity problem
was considered by some as deriving from the non-commutativity character of
operators corresponding to canonical conjugate functions, or just
representing the fact that quantum mechanics, being an essentially
non-classical theory, wouldn't be adjustable to a phase-space description.
Others have thought this problem as irrelevant, for the phase-space
distribution obtained (by Wigner-Moyal-Weyl procedure, for example) will
furnish all the relevant expectation values identical with those of usual
quantum mechanics. Indeed it will give even more, since expectations related
with products of the type $x^np^m$ will be unambiguous calculated within
this formulation of the problem---something the usual approach may do only
after postulating some specific operator construction method.

Others, however, have seen the far reaching consequences of being capable to
construct such a representation, without the inconveniences introduced by
the negativity problem\cite{Bohm,Wiener,Bartlett}.

We will join this last class of authors on their concerns. We have been
arguing, since the first paper of this series\cite{Olavo}, that quantum
mechanics is just a classical statistical theory, defined upon configuration
space, related with the thermodynamic equilibrium situations of physical
systems; this means that it is more restricted than the theory related with
Liouville's equation, which is classical statistical mechanics {\it strictu
sensu}. However, the element of liaison between the Liouville classical
statistical mechanics, defined upon phase-space, and the Schr\"odinger
quantum mechanics upon configuration-space was, until presently, the
phase-space probability density function assuming negative values. If this
function does assume negative values within some ranges of $x$ and $p$
values, then no acceptable physical interpretation may be appended to it.
Then comes the discomfort of having all the theory based upon an element of
irrationality (unphysical).

The task of being capable to obtain positive definite probability densities
will thus fulfill this epistemological void of the theory, giving to it its
final contours. This is the intention of the present paper.

In the second section we will show the method of deriving quantum mechanics,
we had thus far adopted, from classical statistical mechanics\cite{Olavo}.
While doing this, we will show what were our assumptions and why they did
lead to the negativity problem. This critic development of the formalism
will give us insight about the ways we may possibly overcome the
difficulties.

In the third section we will use the insights furnished by the previous
section to derive a method that gives us only positive definite phase-space
probability density functions.

The results of the third section will be used to derive, as a naive
application, the Ehrenfest's theorem, giving the dynamical behavior of the
position and momentum expectations. This will be done in the fourth section.

In the fifth section we will compare the two procedures of finding
phase-space probability density functions. This comparison will be made
based upon the expectation values these two approaches furnish to the
specific example of the harmonic oscillator.

The last section will be devoted to our conclusions.

In the appendix we make a comment about an interesting result, closely
related with our own, made in the literature.

\section{Previous Approach}

In our previous approach to the problem, we began with the phase-space joint
probability density function $F(x,p;t)$ defined as a convolution of two
phase-space probability amplitudes $\phi(x,p;t)$ 
\begin{equation}
\label{1} F(x,p;t)=\int_{-\infty}^{+\infty}\phi^{\dag}(x,2p-p';t)%
\phi(x,p';t)dp'. 
\end{equation}
We then defined the characteristic function $Z_Q(x,\delta x;t)$ as the {\it %
infinitesimal}\ Fourier transform 
\begin{equation}
\label{2} Z_Q(x,\delta x;t)=\int e^{ip\delta x/\hbar}F(x,p;t)dp, 
\end{equation}
and applied the above defined Fourier transform to Liouville's equation,
which we assumed the function $F(x,p;t)$ satisfies, to find the equation
satisfied by $Z_Q$ as 
\begin{equation}
\label{3} -i\hbar\frac{\partial Z_Q(x,\delta x;t)}{\partial t}- \frac{\hbar^2
}{m}\frac{\partial^2 Z_Q(x,\delta x;t)} {\partial x\partial (\delta x)}+
\frac{\partial V(x)}{\partial x}\delta x Z_Q(x,\delta x;t)=0. 
\end{equation}

Because of expressions (\ref{1}) and {\ref{2}), the characteristic function
was given by 
\begin{equation}
\label{4} Z_Q(x,\delta x;t)=\psi^{\dag}(x-\delta x/2;t)\psi(x+\delta x/2;t) 
\end{equation}
where we wrote, for the {\it complex} amplitudes, 
\begin{equation}
\label{5} \psi(x;t)=R(x;t)e^{iS(x;t)/\hbar}, 
\end{equation}
with $R(x;t)$ and $S(x;t)$ real functions. The characteristic function
became, in terms of the $R(x;t)$ and $S(x;t)$ functions 
$$
Z_Q(x,\delta x;t)=\left\{ R(x;t)^2+\left(\frac{\delta x}2\right)^2 \times
\right. 
$$
\begin{equation}
\label{5a} \times \left. \left[ R(x;t)\frac{\partial^2 R(x;t)}{\partial x^2}%
-\left(\frac{\partial R(x;t)} {\partial x}\right)^2\right]\right\}exp\left(
\frac{i}{\hbar} \delta x\frac{\partial S(x;t)}{\partial x}\right), 
\end{equation}
up to second order in the variable $\delta x$, since it is assumed as
infinitesimal and equation (\ref{3}) has to be considered up to order one on
this variable. }

Substituting these expressions in equation (\ref{3}) we got the two
equations 
\begin{equation}
\label{6} \frac{\partial R(x;t)^2}{\partial t}+\frac{\partial}{\partial x}
\left[\frac{R(x;t)^2}{m}\frac{\partial S(x;t)}{\partial x}\right]=0 
\end{equation}
and 
\begin{equation}
\label{7} \frac{-i\delta x}{\hbar}\frac{\partial}{\partial x} \left\{\frac{%
\partial S(x;t)}{\partial t}+\frac 1{2m}\left(\frac{\partial S(x;t)}{%
\partial x}\right)^2+V(x)-\frac{\hbar^2}{2mR(x;t)} \frac{\partial^2 R(x;t)}{%
\partial x^2}\right\}=0, 
\end{equation}
where we used the independence between $x$ and $\delta x$, and also the
infinitesimal character of $\delta x$ to retain only up to power one on this
variable.

These two equations are mathematically equivalent to the Schr\"odinger
equation 
\begin{equation}
\label{8} -\frac{\hbar^2}{2m}\frac{\partial^2 \psi(x;t)}{\partial x^2}+
V(x)\psi(x;t)=i\hbar\frac{\partial \psi(x;t)}{\partial t}, 
\end{equation}
when we substitute the expression (\ref{5}) on it, and separate the real and
imaginary parts.

This process allowed us to find, upon inversion of the Fourier
transformation (\ref{2}), a phase-space probability density which will
simulate all the results of ordinary quantum mechanics (the statistical
moments, such as expectations, square deviations, etc..). However, by the
very definition of the phase-space probability density $F(x,p;t)$ given in (%
\ref{1}), we shall not expect to find positive-definite functions, since the
phase-space amplitudes may oscillate---mainly for the excited states, where
their corresponding configuration-space amplitudes show such an oscillatory
behavior---and their product is defined at different momentum-space points
in (\ref{1}), as required for a convolution.

The process of operator formation was then simply introduced. Since $Z_Q$ is
a characteristic function (in momentum space) of the phase-space probability
density function, all statistical moments may be obtained as 
\begin{equation}
\label{8a} \overline{(x^np^m)}=\int \lim_{\delta x\rightarrow 0} x^n
\left(-i\hbar\frac{\partial}{\partial (\delta x)}\right)^m Z(x,\delta x;t)
dx, 
\end{equation}
where no ambiguity appears, since $x$ and $\delta x$ are independent
variables. Using the decomposition (\ref{4}) we may connect these
expectation values, calculated with the characteristic function $Z_Q$, with
those calculated using the probability amplitudes. This can be accomplished
by means of the expression (\ref{8a}) above with (\ref{4}) for the
characteristic function, after performing an expansion of (\ref{4}) with
respect to the infinitesimal variable $\delta x$. In this case, the position
and momentum operators become $\hat{x}=x$ and $\hat{p}=i\hbar\partial/%
\partial x$, and no longer commute.

We have argued that it was the process of inversion of the infinitesimal
Fourier transformation (\ref{2}) which introduced such negativity into our
phase-space distributions. Now, we may say that it is mainly the imposition
of having the characteristic function $Z_Q$ written as a product that has
introduced this negativity, since it forces the phase-space probability
density to be written as a convolution in momentum space. We note, however,
that, except for this unacceptable behavior---which has no physical
interpretation---the phase-space densities thus derived was really adequate
to the problem, for the reasons above mentioned. We just note that, when
writing the characteristic function as the Fourier integral in (\ref{2}) and
assuming the decomposition in (\ref{4}) we have also introduced an specific
operator construction method\cite{Olavo9}, which will give us a procedure to
evaluating expectation values of functions as $x^np^m$. This means that,
when changing some of these definitions (e.g. the convolution in (\ref{1})),
the operator construction method may also vary, giving possibly new
expectation values for these product. However, the expectation values of the
`lateral functions' $x^n$ and $p^m$ will always have to agree with those
obtained by usual quantum mechanics techniques.

We will now present a method to overcome the negativity problem. In the next
section we will present a new derivation of the problem in which we assume,
from the very beginning the positive character of the phase-space
probability density function.

\section{Positive Densities}

We begin by defining the phase-space probability density as 
\begin{equation}
\label{9} F(x,p;t)=\phi^{\dag}(x,p;t)\phi(x,p;t), 
\end{equation}
where $\phi(x,p;t)$ are probability amplitudes defined upon phase-space.
This function will thus be obviously positive definite.

The characteristic function continues to be defined by expression (\ref{2}),
but now we will also define the {\it characteristic amplitudes} 
\begin{equation}
\label{10}\xi (x^{\prime },x;t)=\int e^{ipx^{\prime }/\hbar }\phi (x,p;t)dp,
\end{equation}
to write $Z_Q$, in terms of these functions, as 
\begin{equation}
\label{11}Z_Q(x,\delta x;t)=\int \xi ^{*}(x^{\prime },x;t)\xi (x^{\prime
}+\delta x,x;t)dx^{\prime },
\end{equation}
which now is a convolution. This was readily expected, since now we are
defining the probability density $F(x,p;t)$ as a product upon phase-space,
and $F(x,p;t)$ and $Z_Q(x,\delta x;t)$ are the Fourier transform of each
other. We thus have fixed the positivity of the phase-space probability
density function while letting the characteristic function to possibly
assume negative values, since this will not be physically unacceptable.

Another way of writing expression (\ref{11}) is 
\begin{equation}
\label{12} Z_Q(x,\delta x;t)=\int \xi^*(x',x;t)e^{i\delta x \hat{p}%
'/\hbar} \xi(x',x;t)dx', 
\end{equation}
where 
\begin{equation}
\label{13} \hat{p}'=-i\hbar\frac{\partial}{\partial x'}. 
\end{equation}

We may substitute expression (\ref{12}) into equation (\ref{3}) to find the
equation satisfied by the characteristic amplitudes $\xi$ as 
\begin{equation}
\label{14} -i\hbar\frac{\partial \xi(x',x;t)}{\partial t}- \frac{%
\hbar^2}{m}\frac{\partial^2 \xi(x',x;t)} {\partial x\partial
x'}+\frac{\partial V(x)}{\partial x}x'\xi(x',x;t)=0, 
\end{equation}
which was expected, since the phase-space amplitudes also satisfy a
Liouville equation\cite{Wiener} and the definition of the $\xi$'s parallels
that of the characteristic function $Z_Q$.

It is obvious that all the statistical moments related to momentum or
position values are still obtainable from the characteristic function $Z_Q$.
In fact any function of the type $O[x,p]$ will have its expectation value,
when calculated using the characteristic amplitudes, given by 
\begin{equation}
\label{15} \overline{O[x,p]}=\int\int \xi^*(x',x;t)\hat{O}[\hat{x},
\hat{p}'] \xi(x',x;t)dx'dx, 
\end{equation}
where 
\begin{equation}
\label{16} \hat{O}[\hat{x},\hat{p}']=O\left[x,-i\hbar\frac{\partial}
{\partial x'}\right] 
\end{equation}
and no ambiguities appear, for $x$ and $x'$ are independent
variables. These results come from the very application of the definitions (%
\ref{8a}) to the new definition (\ref{11}). The explicit results are 
\begin{equation}
\label{17} \overline{(x^np^m)}=\int \int \xi(x',x;t)x^n\left(-i\hbar
\frac{\partial} {\partial x'}\right)^m \xi(x',x;t)
dxdx'. 
\end{equation}

We may now put 
\begin{equation}
\label{18}\xi (x^{\prime },x;t)=\psi (x^{\prime };t)\psi ^{\dag }(x;t)
\end{equation}
to get, for the characteristic function 
\begin{equation}
\label{19}Z_Q(x,\delta x;t)=\psi (x;t)\psi ^{\dag }(x;t)\int \psi ^{\dag
}(x^{\prime };t)exp\left( \delta x\frac \partial {\partial x^{\prime
}}\right) \psi (x^{\prime };t)dx^{\prime },
\end{equation}
giving, for the lateral statistical moments 
\begin{equation}
\label{20}\overline{(p^m)}=\int \psi ^{\dag }(x^{\prime };t)\left( -i\hbar
\frac \partial {\partial x^{\prime }}\right) ^m\psi (x^{\prime
};t)dx^{\prime }
\end{equation}
and 
\begin{equation}
\label{21}\overline{(x^n)}=\int \psi ^{\dag }(x;t)(x^n)\psi (x;t)dx,
\end{equation}
whereas, for the general product, we have 
\begin{equation}
\label{22}\overline{(x^np^m)}=\int \psi ^{\dag }(x;t)(x^n)\psi (x;t)dx\int
\psi ^{\dag }(x^{\prime };t)\left( -i\hbar \frac \partial {\partial
x^{\prime }}\right) ^n\psi (x^{\prime };t)dx^{\prime }.
\end{equation}
Note that, now, the characteristic function became a quadrilinear function
of the amplitudes\cite{Bohm}. This new definition of the characteristic
function will lead to an operator construction method distinct from the one
presented in the last section.

Until now we have not specified which equation the probability amplitudes $%
\psi (x;t)$ satisfy---indeed, this was done in the previous section but,
since our definitions has changed, we are not certain if these amplitudes
still obey the Schr\"odinger equation. However, if we turn our attention to
the operator formation procedure given in (\ref{20}) and (\ref{21}), we thus
see that the expectation of the energy function, given by 
\begin{equation}
\label{22a}\overline{E}=\int \int \left( \frac{p^2}{2m}+V(x)\right)
F(x,p;t)dxdp,
\end{equation}
may be written, in terms of the probability amplitudes, as 
\begin{equation}
\label{22b}\overline{E}=\int \psi ^{\dag }(x;t)\left[ \frac 1{2m}\left(
-i\hbar \frac \partial {\partial x}\right) ^2+V(x)\right] \psi (x;t)dx,
\end{equation}
which implies the Schr\"odinger equation 
\begin{equation}
\label{22c}\frac{\hat p^2}{2m}\psi (x,t)+V(x)\psi (x;t)=E\psi (x;t),
\end{equation}
with $\hat p=-i\hbar \partial /\partial x$. This suffices to prove that the
amplitudes $\psi (x;t)$ still obey the Schr\"odinger equation.

The last expression for the expectation values of coordinate-momentum powers
(\ref{22}) will thus strongly differ from the one we have obtained in the
last section. We also note that, in the present procedure, we still do not
introduce any commutation problem, for $x$ and $x'$ continue to be
independent variables, just like $x$ and $\delta x$ are. Since we showed
that the probability amplitudes satisfy the Schr\"odinger equation, all the
`lateral' expectation values predicted by quantum mechanics will be
reproduced by the present theory. The discrepancy between this section
approach and the one described in the previous section may not be solved
within the realm of the usual quantum formalism, for both approaches predict
the same lateral statistical expectations ($x^n$ or $p^m$ alone), which are
the only ones guaranteed by the above mentioned formalism---the
coordinate-momentum product expectations depending on the particular
operator construction method (the correspondence rule) each approach
proposes.

Using now the definition (\ref{12}) and the decomposition (\ref{18}), we
have 
\begin{equation}
\label{23} \psi^{\dag}(x;t)\psi(x';t)=\int
e^{ipx'/\hbar}\phi(x,p;t)dp, 
\end{equation}
which implies 
\begin{equation}
\label{24} \phi(x,p;t)=\psi^{\dag}(x;t)\varphi(p;t), 
\end{equation}
where 
\begin{equation}
\label{25} \varphi(p;t)=\int
e^{-ipx'/\hbar}\psi(x';t)dx'. 
\end{equation}

With the result (\ref{24}), for the phase-space probability amplitudes, and
the initial imposition (\ref{9}) we get 
\begin{equation}
\label{26} F_n(x,p;t)=|\psi_n(x;t)|^2|\varphi_n(p;t)|^2, 
\end{equation}
as the desired positive definite phase-space probability density function.
This result reinforces the $F_n$ property of giving all the lateral
statistical expectations as the ones predicted by usual quantum mechanics.

In the literature there are some restrictions to the density obtained in (%
\ref{26}). The nature of these restrictions\cite{Cohen} and the explanation
of why we have been successful in overcome them with our present approach
may be found in the appendix.

\section{Ehrenfest's Theorem}

From the above relations it is possible to derive the Ehrenfest's theorem in
a straightforward manner. To see this consider first the equation satisfied
by the characteristic function $Z_Q$ (\ref{3}). Writing this characteristic
function up to second order in the infinitesimal displacement $\delta x$ we
get 
$$
Z_Q(x,\delta x;t)=\psi^{\dag}(x;t)\psi(x;t)\left[1+ \delta x \int
\psi^{\dag}(x';t)\frac{\partial \psi(x';t)}{\partial
x'}dx'+ \right. 
$$
\begin{equation}
\label{26a} \left.+\frac{(\delta x)^2}{2}\int \psi^{\dag}(x';t) 
\frac{\partial^2 \psi(x';t)}{\partial x'^2}%
dx'\right] . 
\end{equation}
Now, taking this expression into (\ref{3}) and separating the zeroth and
first order terms in the infinitesimal displacement $\delta x$, we get the
two equations 
\begin{equation}
\label{26b} -i\hbar\frac{\partial}{\partial t}(\psi^{\dag}(x;t)\psi(x;t))- 
\frac{\hbar^2}{m}\frac{\partial}{\partial x}(\psi^{\dag}(x;t)\psi(x;t)) \int
\psi^{\dag}(x';t)\frac{\partial \psi(x';t)}{\partial
x'}dx'=0, 
\end{equation}
and 
$$
-i\hbar\frac{\partial}{\partial t}(\psi^{\dag}(x;t)\psi(x;t)) \int
\psi^{\dag}(x';t)\frac{\partial \psi(x';t)}{\partial
x'}dx'-
$$
$$
-i\hbar (\psi^{\dag}(x;t)\psi(x;t)) \int
\psi^{\dag}(x';t)\frac{\partial \psi(x';t)}{\partial
x'}dx'- 
$$
\begin{equation}
\label{26c} -\frac{\hbar^2}{m}\frac{\partial}{\partial x}(\psi^{\dag}(x;t)%
\psi(x;t)) \int \psi^{\dag}(x';t)\frac{\partial^2 \psi(x';t)
}{\partial x'^2}dx'+\frac{\partial V(x)}{\partial x}%
(\psi^{\dag}(x;t)\psi(x;t))=0. 
\end{equation}

Multiplying the first equation by $x$ and integrating we find 
\begin{equation}
\label{26d} \frac{\partial \overline{x}}{\partial t}=\overline{p}, 
\end{equation}
while, integrating the second equation in $x$, we find 
\begin{equation}
\label{26e} \frac{\partial \overline{p}}{\partial t}=- \overline{\left(\frac{%
\partial V(x)}{\partial x}\right)}. 
\end{equation}
These last two equations are exactly the mathematical expressions of the
Ehrenfest's theorem\cite{Schiff}.

Moreover, with the probability density on configuration space written as $%
\rho(x;t)=|\psi(x;t)|^2$, the equation (\ref{26b}) becomes 
\begin{equation}
\label{26f} \frac{\partial \rho(x;t)}{\partial t}+\frac{\partial}{\partial x}
\left[\frac{\overline{p(t)}}{m}\rho(x;t)\right]=0, 
\end{equation}
while using this equation into (\ref{26c}) we get 
\begin{equation}
\label{26g} \rho(x;t)\left[\frac{\partial \overline{p(t)}}{\partial t}+ 
\frac{\partial V(x)}{\partial x}\right]+\frac{\overline{(\delta p)^2}}{m} 
\frac{\partial \rho(x;t)}{\partial x}=0, 
\end{equation}
where
\begin{equation}
\label{26h}
\overline{(\delta p)^2}=\overline{p(t)^2}-\overline{p(t)}^2.
\end{equation}

These last two equations are precisely those we would get if we integrate the
Liouville equation in the variable $p$ after multiplying it by $1$ and $p$,
respectively, using a separable density function of the type given in (\ref
{26})\cite{Olavo4}.

\section{Application: Harmonic Oscillator}

We now want to make a simple application of the previous results to the
harmonic oscillator problem. Thus, we begin with the harmonic oscillator
probability amplitude given by 
\begin{equation}
\label{27} \psi_n(x;t)=\left(\alpha^2/4\right)^{1/2}\left(\frac{1}{2^n n!}%
\right) ^{1/2}e^{-\alpha^2 x^2/2}H_n(\alpha x)e^{-iE_n t/\hbar}, 
\end{equation}
where $H_n(x)$ are the Hermite polynomials and $\alpha=\sqrt{m\omega/\hbar}$%
, with $m$ the mass and $\omega$ the frequency related with the problem.

Using the approach of the second section, the phase-space probability
density function may be written as\cite{Groenewold,Takabayasi,Dahl,Hillary} 
\begin{equation}
\label{28} F_n^B(x,p;t)=\frac{1}{\pi\hbar}(-1)^n exp\left(-\frac{2H}{%
\hbar\omega} \right)L_n\left(\frac{4H}{\hbar\omega}\right), 
\end{equation}
which clearly assumes, for the excited states, negative values within some
ranges of the variable $H=p^2/2m+m\omega^2x^2/2$.

The phase-space probability density function of the present approach we
uphold in the third section is simply 
\begin{equation}
\label{29} F_n^A(x,p;t)=|\psi_n(x;t)|^2|\varphi_n(p;t)|^2, 
\end{equation}
where $\varphi_n(p;t)$ is the solution of the problem on the momentum
space---the Fourier transform of $\psi(x;t)$.

The results obtained for the expectation values by these two phase-space
probability densities are exhibited in Table I. We may see from this table
that the expectations will coincide for all cases but those involving the
product $x^np^m$. While the phase-space probability density function (\ref
{28}) presents some correlation between the variables $x$ and $p$, these
correlations are absent from the one in (\ref{29}). In the present harmonic
oscillator case, this will affect, for example, the expectations of the
energy dispersions, which become higher for the uncorrelated density.

This result was expected since, for the excited states, the densities spread
more and more with respect to the variable $H$ and its dispersion,
calculated with the probability density $F^B$, remains constant only because
this density assumes negative values within some ranges of $H$; since this
negativity regions do not appear in the probability density $F^A$, we may
expect the spread in the energy to get larger, following the density spread
itself. Thus, the probability density $F^B$ presents the strange behavior of
having the momentum and position dispersion getting larger and larger, while
the energy dispersion being kept constant.

We have also plotted the first three phase-space probability density
functions according to the prescription (\ref{29}). They are given by
figures I, II and III. These figures may be contrasted with those related
with the prescription (\ref{28}) (see, for example, Olavo\cite{Olavo9}).

\section{conclusion}

In this paper we have been successful in obtaining positive definite
phase-space probability density functions. Although this result does not
change very deeply the formal aspects of the quantum theory, since this
theory will continue to be given by the Schr\"odinger equation, it does
constitute an important epistemological acquisition. Indeed, since the very
beginning of this series we have being arguing that quantum mechanics is a
mere classical statistical mechanics, performed upon configuration space,
for systems in thermodynamic equilibrium\cite{Olavo4}. However, it will be
rather inconvenient for such an epistemology to have its main symbol of
liaison (the phase-space probability density) between these two levels of
descriptions (phase-space and configuration space) not being capable of
having a physical interpretation.

Thus, being capable of calculating the probability density functions on
phase-space not suffering from the negativity problem gives us the final
justification to our previous assertions. Now the theory has all its symbols
with some acceptable physical interpretation attached to it and we may
finally say, without fearness, that quantum mechanics is just an {\it %
ensemble} statistical theory performed upon configuration space and related
with thermodynamic equilibrium situations.

The next paper will be devoted to the epistemological consequences of the
formalism derived in this series of papers\cite{Olavo} and will constitute
its natural logical closure.

\appendix

\section{Restriction upon the Density}

Cohen\cite{Cohen} has derived a generalized method of obtaining phase-space
density functions. He was aiming at giving a general procedure of obtaining
all the phase-space probability distributions proposed at the time his paper
was published. He thus begins by writing this phase-space probability
density function as 
$$
F(x,p;t;f)=\frac{1}{4\pi^2}\int\int\int e^{-i\theta x-i\tau
p+i\theta u} f(\theta,\tau;t)\times
$$
\begin{equation}
\label{a1}
\times\psi^{\dag}(u-\frac 12\tau\hbar;t)\psi(u+\frac
12\tau\hbar;t) d\theta d\tau du. 
\end{equation}

In this case, a wide variety of probability densities may be obtained by
just fixing the function $f(\theta,\tau;t)$ (e.g. the Wigner function in (%
\ref{2}) may be found by fixing $f=1$).

In his conclusions he argues that: ``It is commonly held that the
uncertainty principle by itself precludes the possibility of the existence
of a joint distribution of position and momentum. However, this is not so.
For example, the choice 
\begin{equation}
\label{a2} f(\theta,\tau;t)=\frac{\int\int |\psi(x)|^2|\phi(p)|^2 e^{i\theta
x +i\tau p}dxdp}{\int \psi^{\dag}(u-\frac 12\tau \hbar)e^{i\theta u}
\psi(u+\frac 12\tau \hbar)du} 
\end{equation}
leads to 
\begin{equation}
\label{a3} F(x,p;t)=|\psi(x)|^2|\phi(p)|^2, 
\end{equation}
which is certainly a well-defined joint distribution and from which the
uncertainty principle follows in the usual manner. The reason why a true
joint distribution cannot be defined is because no choice of $f$ yields a
distribution which gives the correct quantum mechanical expectation values
for all observables when calculated through phase-space integration. That
is, no $f$ exists such that, if the correspondence of quantum to classical
variables (..) is 
\begin{equation}
\label{a4} g(x,p)\rightarrow \hat{G}, 
\end{equation}
for some $f$, then also 
\begin{equation}
\label{a5} H(g(x,p))\rightarrow H(\hat{G}), 
\end{equation}
for the same $f$, where $H$ is any function''.

However, this was exactly what we have obtained in the third section of this
paper. Indeed, our method of operator formation was 
\begin{equation}
\label{a6} \hat{x}=x \mbox{ and } \hat{p}=-i\hbar\frac{\partial}{\partial
x'} 
\end{equation}
for which any function $H$ will be such that 
\begin{equation}
\label{a7} H(g(x,p))\rightarrow H(\hat{G})=H(g(\hat{x},\hat{p})), 
\end{equation}
since $\hat{x}$ and $\hat{p}$ commute, denying ambiguities related with
operator ordering to appear.

This is precisely the point of disagreement. When constructing his
correspondence rule, Cohen\cite{Cohen2} has explicitly assumed that the
operators related with the position and the momentum are those given by $%
\hat x=x$ and $\hat p=-i\hbar \partial /\partial x$, which do not commute.
This precludes, in his approach, the possibility of finding a function $f$
which gives, at the same time, a specific correspondence rule, the density
as in (\ref{a3}) (for example), and all the expectations values predicted by
quantum mechanics as equal to those calculated by using the derived
correspondence rule.

Thus, with the choice (\ref{a1}) he was able to find a function $f$ which
gives the phase-space probability density as (\ref{a3}). With this function $%
f$ he was able to find the correspondence rule which takes a function $g(x,p)
$ into its related operator $\hat G$. But when calculating the expectation
values with the phase-space probability density and the operator $\hat G$,
he was not able to make the results agree. Looking at our own development,
we may see that the problem is due to the fact that, in his last derivation
step, Cohen\cite{Cohen2} imposes that the variables $x$ and $p$ be
substituted by the non-commuting operators $\hat x=x$ and $\hat p=-i\hbar
\partial /\partial x$, which may not be accommodated within our present
approach. Indeed, it is the commuting of the operators in (\ref{a6}) that
guarantees the validity of the expectation values calculated with the
uncorrelated probability density (\ref{a3}), as we can see by (\ref{22}),
and it is obvious that we will have exactly the same results while
calculating the expectations with the phase-space probability density $F^A$
or using the operator given by (\ref{a7}), which is our correspondence rule.

\newpage
\begin{center}
\Huge
\mbox{TABLES}
\end{center}
\normalsize
\vspace{30pt}

\begin{table}
 \begin{center}
  \begin{tabular}{|c|c|c|c|c|c|c|c|c|} \hline
{$n$}&$ \overline{E}  $       &
      $ \overline{x^2}$       & 
      $ (\Delta x\Delta p)$   & 
      $ (\overline{x^4})$     & 
      $ (\overline{x^2p^2})_A$& $ (\overline{x^2p^2})_B$ &
      $ (\Delta E)_A^2$       & $ (\Delta E)_B^2       $  \\ \hline \hline
 0 & 1/2 & 1/2 & 1/2 &   3/4 &  1/4 &  1/4 &  1/4 & 1/4 \\ \hline
 1 & 3/2 & 3/2 & 3/2 &  15/4 &  9/4 &  5/4 &  3/4 & 1/4 \\ \hline
 2 & 5/2 & 5/2 & 5/2 &  39/4 & 25/4 & 13/4 &  7/4 & 1/4 \\ \hline
 3 & 7/2 & 7/2 & 7/2 &  75/4 & 49/4 & 25/4 & 13/4 & 1/4 \\ \hline  
  \end{tabular}                        
\end{center}
 \caption{Some expectation values calculated with the present probability density
function in phase space ($A\equiv F^A$) and the one usually found in the 
literature ($B\equiv F^B$). The results coincide whenever the index is absent.
\label{tabela1}}
\end{table}

\newpage

\begin{center}
\Huge
\mbox{FIGURES}
\end{center}
\normalsize

\vspace{30pt}

\mbox{Figure 1 - Plot of the phase space probability density function for
N=0.}

\vspace{30pt}

\mbox{Figure 2 - Plot of the phase space probability density function for
N=1.}

\vspace{30pt}

\mbox{Figure 3 - Plot of the phase space probability density function for
N=2.}


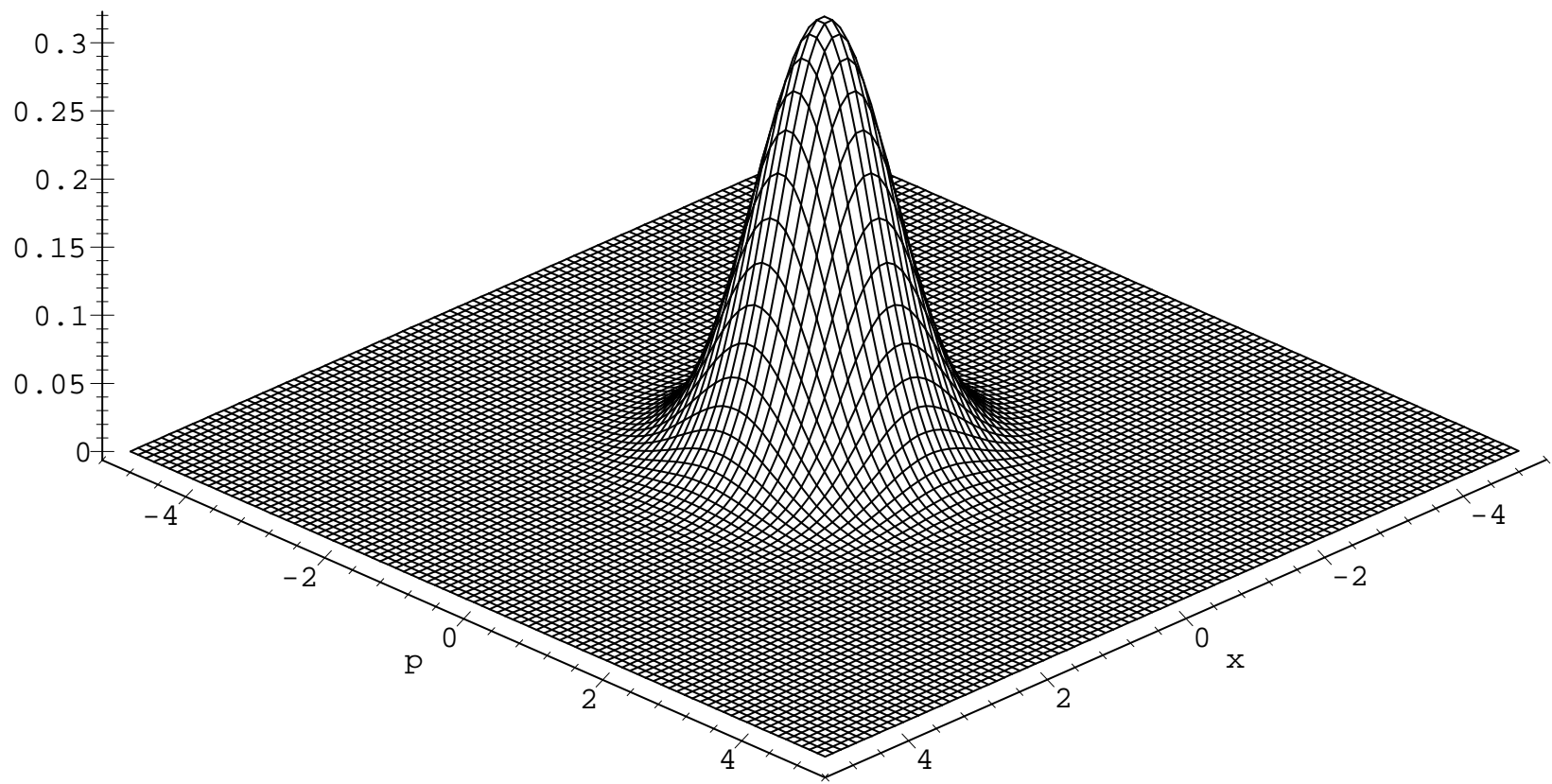
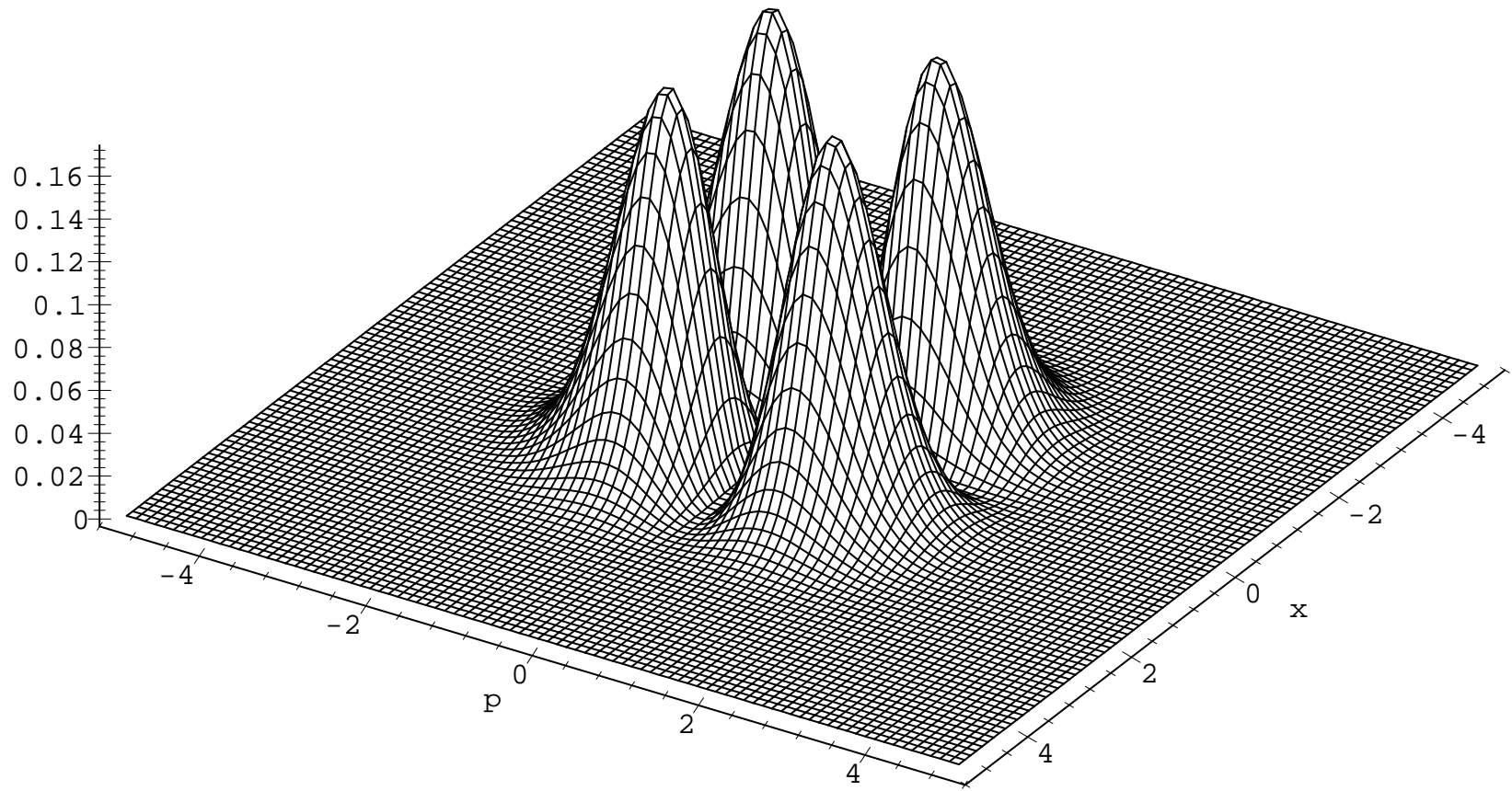
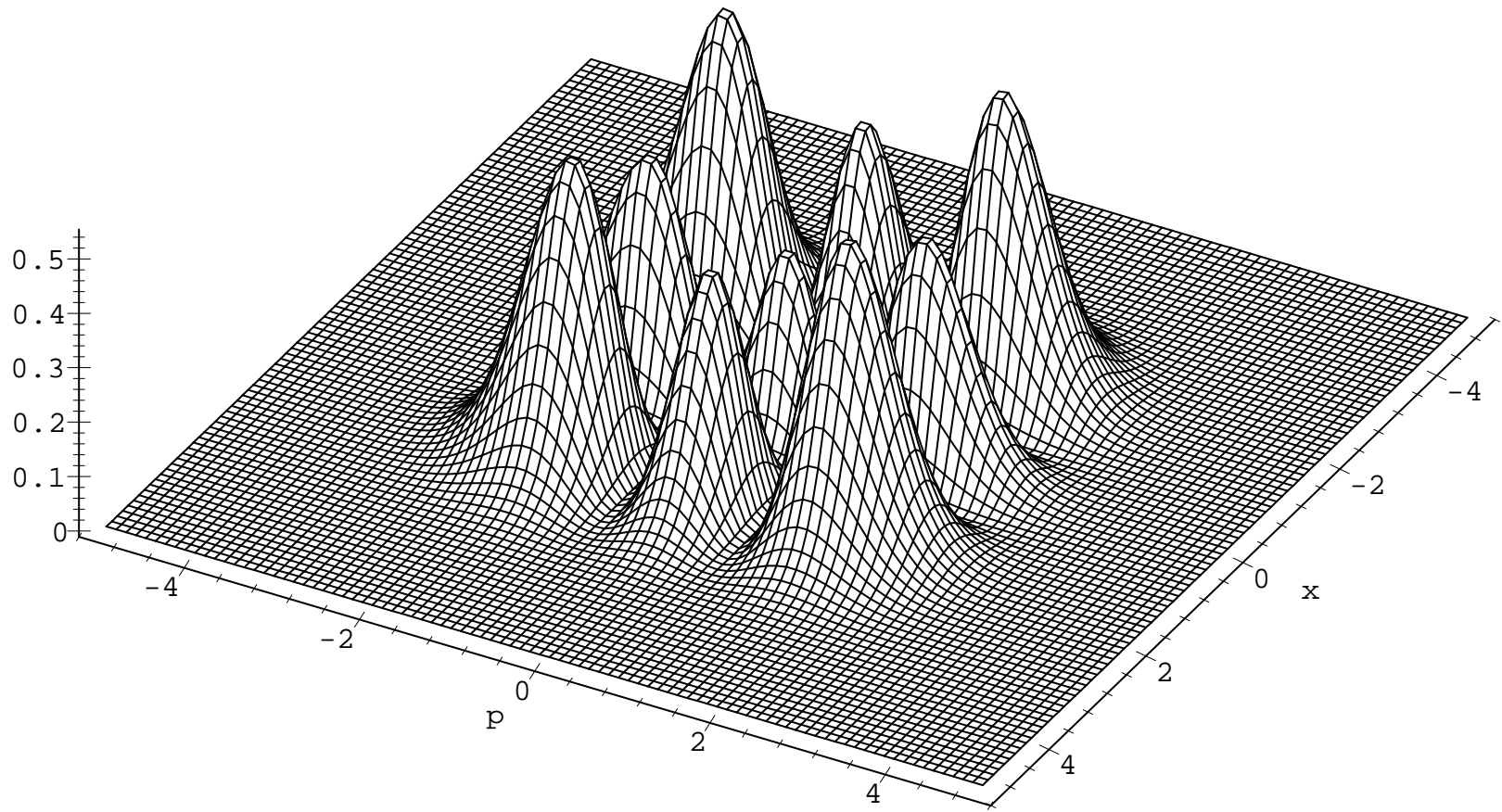
\end{document}